\documentclass[letter]{aa}  

\usepackage{graphicx}
\usepackage[varg]{txfonts}
\usepackage[flushleft]{threeparttable}
\usepackage[colorlinks=true,allcolors=blue]{hyperref}
\begin{document} 

   \title{Making the unmodulated pyramid wavefront sensor smart}
   \subtitle{II. First on-sky demonstration of extreme adaptive optics with deep learning}

   \titlerunning{First on-sky demonstration of extreme adaptive optics with deep learning}
   \authorrunning{R. Landman et al.}

   \author{
   R. Landman\inst{1}
   \and
   S.~Y.~Haffert\inst{1,2}
    \and
    J. D. Long\inst{3}
   \and 
   J.~R.~Males\inst{2}
    \and
   L.~M.~Close\inst{2}
   \and
   W.~B.~Foster\inst{2}
   \and
   K.~Van Gorkom\inst{2}
   \and
   O. Guyon\inst{2,4,5,6}
    \and
   A.~D.~Hedglen\inst{7}
   \and
   P.~T.~Johnson\inst{2}
   \and
   M.~Y.~Kautz\inst{4}
   \and
   J.~K.~Kueny\inst{4}
   \and
   J.~Li\inst{2}
   \and
   J.~Liberman\inst{4}
    \and
   J.~Lumbres \inst{4}
   \and
   E.~A.~McEwen \inst{4}
   \and
   A.~McLeod \inst{8}
   \and
   L. Schatz\inst{9}
   \and
   E.~Tonucci\inst{1}
   \and
   K.~Twitchell\inst{4}
   }

   \institute{Leiden Observatory, Leiden University, PO Box 9513, 2300 RA Leiden, The Netherlands \\
              \email{rlandman@strw.leidenuniv.nl}
    \and
    Steward Observatory, The Unversity of Arizona, 933 North Cherry Avenue, Tucson, Arizona
    \and
    Center for Computational Astrophysics, Flatiron Institute, 162 5th Avenue, New York, New York
    \and
    Wyant College of Optical Sciences, The University of Arizona, 1630 E University Blvd, Tucson, Arizona
    \and
    Subaru Telescope, National Observatory of Japan, National Institutes of Natural Sciences, 650 N. A'ohoku Place, Hilo, Hawai'i
    \and
    Astrobiology Center, National Institutes of Natural Sciences, 2-21-1 Osawa, Mitaka, Tokyo, Japan
    \and
    Northrop Grumman Corporation, 600 South Hicks Road, Rolling Meadows, Illinois
    \and
    Draper Laboratory, 555 Technology Square, Cambridge, Massachusetts
    \and
     Starfire Optical Range, Kirtland Air Force Base, Albuquerque, New Mexico
    }

   \date{Received ; accepted}

 
  \abstract 
  %
   {Pyramid wavefront sensors (PWFSs) are the preferred choice for current and future extreme adaptive optics (XAO) systems. Almost all instruments use the PWFS in its modulated form to mitigate its limited linearity range. However, this modulation comes at the cost of a reduction in sensitivity, a blindness to petal-piston modes, and a limit to the sensor's ability to operate at high speeds. Therefore, there is strong interest to use the PWFS without modulation, which can be enabled with nonlinear reconstructors. Here, we present the first on-sky demonstration of XAO with an unmodulated PWFS using a nonlinear reconstructor based on convolutional neural networks. We discuss the real-time implementation on the Magellan Adaptive Optics eXtreme (MagAO-X) instrument using the optimized TensorRT framework and show that inference is fast enough to run the control loop at >2 kHz frequencies. Our on-sky results demonstrate a successful closed-loop operation using a model calibrated with internal source data that delivers stable and robust correction under varying conditions. Performance analysis reveals that our smart PWFS achieves nearly the same Strehl ratio as the highly optimized modulated PWFS under favorable conditions on bright stars. Notably, we observe an improvement in performance on a fainter star under the influence of strong winds. These findings confirm the feasibility of using the PWFS in its unmodulated form and highlight its potential for next-generation instruments. Future efforts will focus on achieving even higher control loop frequencies (>3 kHz), optimizing the calibration procedures, and testing its performance on fainter stars, where more gain is expected for the unmodulated PWFS compared to its modulated counterpart.

}


   \maketitle
%

\begin{table*}[]
\begin{center}
\caption{Targets, conditions, and results of the on-sky tests.}
\label{tab:observations}
\bgroup
\def\arraystretch{1.2}
\begin{tabular}{cccc|cccc|c}
Date       & Target  &\begin{tabular}[c]{@{}c@{}}I-band\\  magnitude\end{tabular} & \begin{tabular}[c]{@{}c@{}}Observing band/ \\  wavelength\end{tabular} & \begin{tabular}[c]{@{}c@{}}Modulation\\  radius\end{tabular} & \begin{tabular}[c]{@{}c@{}}Reconstruction\\ method\end{tabular} & Seeing & \begin{tabular}[c]{@{}c@{}}Loop \\ frequency\end{tabular} & \begin{tabular}[c]{@{}c@{}}Estimated\\ Strehl ratio\end{tabular} \\ \hline
2024-11-16 & $\alpha$ Eri &  0.6 & CH4 &-                                                                   & Neural Network                                                              &    $\sim$0.43"    & 2 kHz                                                     &          58.1\%                                                        \\
2024-11-16 & $\alpha$ Eri & 0.6 & CH4 & 3 $\lambda/D$                                                                & Linear                                                          &    $\sim$0.43"    & 2 kHz                                                     &         62.7\%                                                         \\\hline
2024-11-16 & AF Lep   & $\sim$ 5.7 & z & -                                                                   & Neural Network                                                              &   $\sim$0.60"     & 2 kHz                                                     &              40.1\%                                                    \\
2024-11-16 & AF Lep   & $\sim$ 5.7 & z & 3 $\lambda/D$                                                                  & Linear                                                          &    $\sim$0.60"    & 2 kHz                                                     &          20.6\%                                                        \\\hline
2024-11-16 & $\pi$ Pup    & 0.55 & CH4 &-                                                                   & Neural Network                                                              &    $\sim$0.48"    & 2 kHz                                                     &          48.6\%                                                         \\\hline
2024-11-20 & $\pi$ Pup  & 0.55  & CH4 & -                                                                  & Neural Network                                                              &    $\sim$0.63"    & 2 kHz                                                     & 22.2\%$^{*}$                                                                \\
2024-11-20 & $\pi$ Pup  & 0.55 & CH4 & -                                                                   & Neural Network                                                              &    $\sim$0.63"    & 3.6 kHz                                                   & 27.8\%$^{*,+}$                                              \\
2024-11-20 & $\pi$ Pup  & 0.55 & CH4 & 3 $\lambda/D$                                                                  & Linear                                                          &   $\sim$0.63"     & 2 kHz                                                     & 33.0\%$^{*}$                                                             
\end{tabular}
\egroup
\end{center}
 \begin{tablenotes}
          \small
          \item \textbf{Notes.} The CH4 narrowband filter has a central wavelength of 875 nm and a bandwidth of 26 nm. $^*$The Strehl ratio was estimated on the companion due to saturation of the PSF of the primary. $^+$ Only frames unaffected by the latency issue were selected.
    \end{tablenotes}
\end{table*}

\section{Introduction}\label{sec:introduction}
Extreme adaptive optics (XAO; \citealt{guyon2018_xao_review}) is essential for reaching the diffraction limit of both current 10-meter-class telescopes and next-generation extremely large telescopes (ELTs). Enhancing the performance of these XAO systems is critical for directly imaging Earth-like planets in reflected light using future high-contrast imaging instruments. The majority of current and planned XAO systems use a pyramid wavefront sensor \citep[PWFS;][]{Ragazzoni1996_pwfs} as their primary wavefront sensor \citep[e.g.,][]{pinna2016soul, fitzsimmons2020gpi,kasper2021pcs,males2022magao, Males2024_gmagaox, haffert2024_gmagaox, bond2022_harmoni_ao,  perera2022gpi, boccaletti2022upgrading}. The preference for the PWFS stems from its enhanced sensitivity compared to the Shack-Hartmann wavefront sensor, which enables faster speeds and better performance on fainter stars \citep[e.g.,][]{Ragazzoni1999_pwfs_sensitivity,Correia2020MNRAS_PWFS}.

A major disadvantage of the PWFS is that it has a nonlinear response, limiting the dynamic range that can be obtained with conventional reconstruction techniques \citep[e.g.,][]{Deo2019_optical_gain_tracking, Chambouleyron2020_opticalgain}. To mitigate this, most instruments use modulation to increase its linearity range. However, there are several disadvantages to this modulation. First, it reduces the PWFS's sensitivity, especially to low-order modes, limiting its loop speed and degrading the performance on faint stars \citep{Chambouleyron2023A&A_noise_propagation_ffwfs, Agapito2023_nonmodulated_pwfs}. Second, it hinders its ability to sense petal-piston modes, which are critical for future segmented telescopes \citep{Bertrou-Cantou2022_petalpiston_pwfs,Hedglen2022_segment_phasing_pwfs, Engler2022_flipflop}. Finally, the modulator also puts a limit on the speed at which the control loop can be run on bright stars, as the beam needs to be modulated with at least the same speed as the control loop frequency. 

A promising approach to avoid the need for modulation is the use of nonlinear reconstructors. These algorithms can provide a software-based solution without a reduction in sensitivity. There have been many different attempts at developing nonlinear reconstructors for the PWFS, but their effectiveness has not yet been demonstrated on-sky. A first group of reconstruction algorithms use optical models of the PWFS to perform a nonlinear estimation of the wavefront \citep[e.g.,][]{Hutterer2018_hutterer_landweber, Frazin2018_gradient_based_estimation, Hutterer2023_nonlinear_reconstruction_maths, Chambouleyron2023_gs_reconstruction, Haffert2024_zernike_nonlinear}. These methods rely on an accurate model of the optical system of the PWFS and often need multiple iterations to converge to an accurate reconstruction. A second group of algorithms use neural networks (NNs) as approximate arbitrary functions in an attempt to learn the inverse relation between PWFS measurements and the wavefront \citep{2018SPIE10703E..1FS_swanson_cnn_pred,Landman2020_nonlinear_cnn, Archinuk2023_nonlinear_reconstruction_ml,Wong2023_nn_reconstruction, Pou2024OExprRL_ML_unmodulated,Weinberger2024A&A}. While many works have shown the capability of these data-driven algorithms for nonlinear reconstruction, there have been no on-sky demonstrations so far. The main issues for on-sky XAO tests are the real-time implementation and the large gap between simulations, lab experiments, and on-sky conditions. In \citet[hereafter Paper I]{Landman2024}, we presented closed-loop lab experiments with the Magellan Adaptive Optics eXtreme (MagAO-X) instrument using a convolutional neural network (CNN) reconstructor. 

In this Letter we present the first on-sky demonstration of a nonlinear reconstructor for the unmodulated PWFS of the MagAO-X system. Section \ref{sec:methods} discusses the real-time implementation of the CNN reconstructor within the MagAO-X software infrastructure. Section \ref{sec:results} presents the on-sky results. Finally, Sect. \ref{sec:conclusions} summarizes the results and lists our conclusions.

\begin{figure*}
    \centering
    \includegraphics[width=\linewidth]{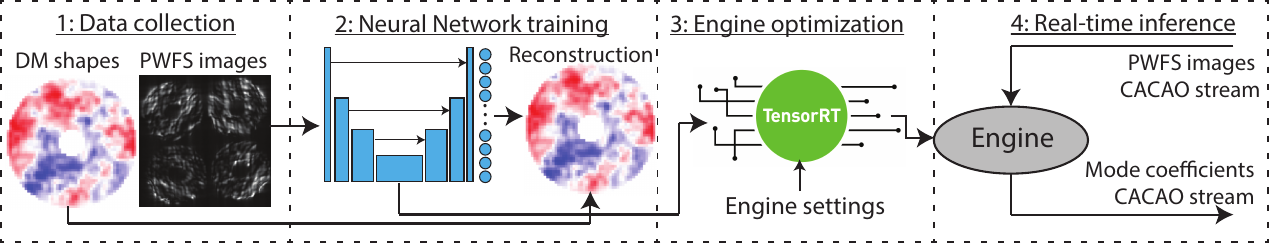}
    \caption{Steps required to run the NN on-sky. First, training data is collected using the internal source by applying random shapes on the DM and recording the resulting PWFS images. After that, the NN is trained on these data and is converted to an optimized TensorRT engine. This engine is loaded in the MagAO-X control software and runs in real time, picking up the PWFS image stream and writing to the mode coefficient stream.}
    \label{fig:pipeline}
\end{figure*}
\section{Methods}\label{sec:methods}
\subsection{MagAO-X}\label{sec:magaox}
MagAO-X is an XAO system that specializes in high-contrast imaging observations at optical wavelengths \citep{males2018magaox, close2018magaox, males2022magao}. It was designed for the 6.5-meter Magellan Clay Telescope at Las Campanas Observatory. MagAO-X uses two optical tables, with the "upper" bench positioned above the "lower" bench. The f/11 beam of the telescope is injected into the instrument at the upper bench. Here the beam goes through several corrective optics: the K mirror that de-rotates the pupil, a tip-tilt mirror that steers the pupil, and the atmospheric dispersion compensator. The upper bench also contains the Alpao-97 deformable mirror (DM) and the Boston Micromachines 2K DM, which work together in a Woofer-Tweeter architecture. The beam is relayed through a periscope systems to the lower optical bench, where most of the science instrumentation is located.

After the periscope system, a dichroic beamsplitter splits the light into two paths: the science channel and the PWFS channel. The beam for the PWFS is spatially filtered with a size of 1.36 arcseconds in diameter. MagAO-X uses a PI stage as the modulator. The (un)modulated beam is focused at f/60 on top the pyramid tip, after which a custom achromatic triplet lens collimates the beam onto an electron-multiplying CCD camera (OCAM2K). The pyramid pupils are separated by 120 pixels and sampled by 112 pixels at the full frame resolution of the OCAM2K. However, the system is always operated in bin2 mode, which creates an effective sampling of 56 pixels across the pupil and 60 pixels of separation between the pyramid pupils. The MagAO-X PWFS is described in more detail in \citet{schatz2018design}. For our tests, we used the I band for wavefront sensing.

The system underwent multiple upgrades since its first on-sky photons \citep{males2024magao}. One of the new additions is a high-order Boston Micromachines 1k DM that sits in the coronagraph arm of the instrument and is not seen by the PWFS \citep{kueny2024magao}. The new kilo DM allows us to apply focal plane wavefront control techniques for non-common path aberration \citep[NCPA;][]{van2021characterizing, kueny2024magao} correction or dark hole digging \citep{haffert2024sky} without disturbing the high-order PWFS loop. This architecture circumvents the optical gain problems that plague systems that need to offload NCPA to their PWFS. The science beam is focused by an off-axis parabola that creates a f/69 beam that is directed onto the science cameras. The science cameras sample the point spread function (PSF) with 3 pixels per $\lambda/D$ at H$\alpha$, which is 5.98 mas pixel$^{-1}$ on-sky \citep{Long_2025}.

\subsection{Model training}
We adopted the same U-net architecture \citep{2015arXiv150504597R_unet} and almost the same training procedure as described in Paper I. For a detailed description of the calibration of the NN, we refer the reader to Paper I. Instead of 1000 modes, we reconstructed and controlled 1563 tweeter modes, which is the standard number of modes used for MagAO-X. We also slightly modified the training procedure. Previously, we trained the CNN directly on the modal coefficients, which were obtained by projecting phase screens spanned by the DM onto the modal basis. This required the inversion of the transformation matrix of the modal basis, and we noticed that the regularization used in this inversion affected the obtained modal coefficients. In this study we instead incorporated the chosen modal basis in the forward model. We optimized the following loss function:
\begin{equation}
    J = \left<\frac{\sqrt{\sum_i (y_{\textrm{true}, i} - y_{\textrm{pred}, i})^2}}{\sqrt{\sum_i y_{\textrm{true}, i}^2}+\epsilon}\right>,
\end{equation}
with
\begin{equation}
    y_{\textrm{pred, i}} = \mathbf{M} \cdot \textrm{CNN}(I),
\end{equation}
where $<>$ denotes the mean over a sampled batch, $y_{\textrm{true,i}}$ the amplitude of DM actuator $i$, $\epsilon$ a small term used to avoid divergence for a very small input RMS, $\textbf{M}$ the projection matrix to go from modal coefficients to DM actuator amplitudes, and $I$ the preprocessed PWFS images. For more details on the loss function, we refer the reader to Paper I. We used $\epsilon=0.005$ in this work and used the default modal basis used in MagAO-X for the CNN, to allow for quick switching between unmodulated and modulated modes and enable a fair comparison. The training data were collected using the internal source at the end of the night before the tests were conducted. During the data collection, we ran a tip-tilt loop with a small gain since we observed some long-term drifts during the data collection in initial tests. Still, small drifts in the reference wavefront due to, for example, temperature variations may impact the results presented here. A new model was trained for the November 16 and 20, 2024, observations.

\subsection{Real-time implementation using TensorRT}\label{sec:real_time}
To run the XAO loop at multiple-kilohertz frequencies, we need a highly optimized implementation of the inference pipeline within the MagAO-X software suite. A visual representation of the steps required is shown in Fig. \ref{fig:pipeline}. After training the model on data from the internal source, we converted the model to an optimized engine using TensorRT\footnote{https://github.com/NVIDIA/TensorRT}, as also suggested in \citet{Pou2024SPIE_real_time_cosmic}. We measured the latency of the TensorRT model using the profiling methods from TensorRT, and the results are reported in Table \ref{tab:latency}. We see that the latency is $<250$ $\mu s$ at single precision and $<125$ $\mu s$ at half precision on the upgraded RTX 4090 GPU present in MagAO-X. This means the engine is fast enough to run at multiple-kilohertz frequencies and is much faster than the values we report for the Python implementation in Paper I, for which we used the old RTX 2080 Ti. We observe that the maximum latency is close to the median value, indicating that the inference time is very stable. We were unfortunately unable to test the half precision model on-sky due to incompatibility with the rest of the software. For reference, the latency of the standard matrix-vector multiplication (MVM) is about $\sim140 $$\mu$s at double precision. In fact, the last linear layer of the CNN, with equivalent size as the standard MVM, is the most computationally expensive step for the CNN reconstructor.

\begin{table}[htbp]
\caption{Profiling results of the TensorRT engine on the RTX 4090 GPU in MagAO-X. The time includes memory transfer to and from the GPU and the inference with the model.}
\begin{center}
\bgroup
\def\arraystretch{1.2}

\label{tab:latency}
\begin{tabular}{l|c|c}
Inference time  & Single precision & Half precision \\ \hline
Median          & 240 $ \mu$s           & 100 $ \mu$s         \\
Min             & 235 $ \mu$s           & 96 $ \mu$s         \\
Max             & 247 $ \mu$s           & 125 $ \mu$s         \\
95\% percentile & 243 $ \mu$s       & 104 $ \mu$s        
\end{tabular}
\egroup
\end{center}
\end{table}

We implemented the reconstruction as a MagAO-X application. We picked up the \textit{aol1\_imWFS2} stream from CACAO \citep{Guyon2018_cacao}, which contains the reference-subtracted and normalized PWFS images. We then split the four PWFS pupils into different channel inputs to the CNN. This preprocessed input was then propagated through the optimized engine using the TensorRT C++ API to obtain the output modal coefficients. These are written to the \textit{aol1\_modevalWFS} stream, which is then picked up again by CACAO for the real-time control. This stream is usually written by the MVM reconstruction, but this was turned off when running the CNN reconstruction.

\section{Results} \label{sec:results}
\begin{figure}[htbp]
    \centering
    \includegraphics[width=\linewidth]{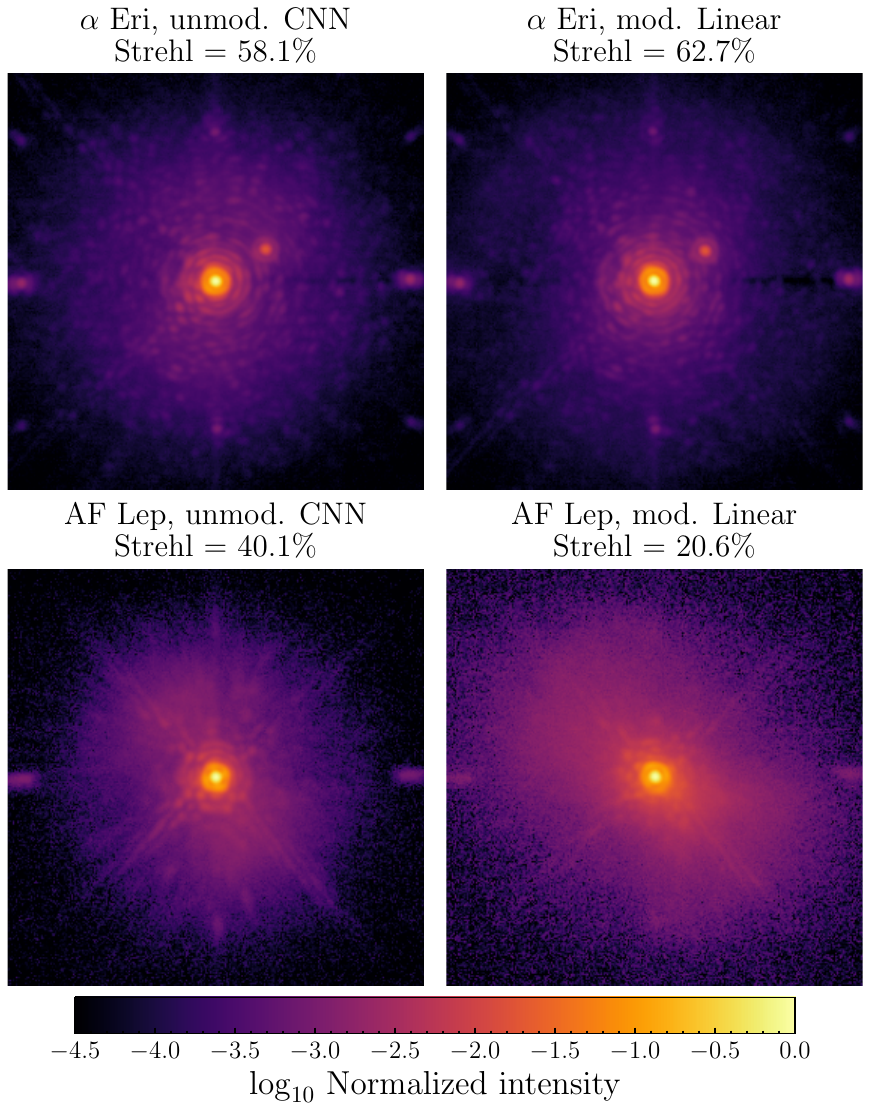}
    \caption{On-sky integrated PSFs for a subset of the tests, comparing the performance between the unmodulated PWFS with the CNN reconstructor and the standard MagAO-X operation using a linear reconstructor and 3 $\lambda/D$ modulation. The estimated Strehl ratio for each of these observations is noted above the image.}
    \label{fig:psfs}
\end{figure}

We tested the performance of our CNN reconstructor under different conditions and on different targets. An overview of the observations is presented in Table \ref{tab:observations}. Before these tests, we first closed the loop and tuned the modal gains using the conventional linear reconstructor with the modulated PWFS. Subsequently, the modulator was turned off and we switched to the CNN reconstructor while keeping the same operational settings. This switching was then repeated to ensure no significant changes in conditions occurred between tests. This should enable a relatively fair comparison of the two methods. The on-sky Strehl ratio was estimated in the following way: First, we simulated a diffraction-limited PSF with the same spatial sampling as the obtained data. We then calculated the ratio between the encircled energy within a central aperture of 1.22 $\lambda/D$ in radius and between 1.22 and 30 $\lambda/D$. This measurement was repeated for the on-sky data, and the on-sky Strehl ratio was estimated by calculating the ratio of these values between the on-sky data and the simulated diffraction-limited image. For the observations of binary stars, we first masked the companion star in the on-sky data before calculating the reference intensity. For some of the on-sky observations, the central region of the PSF is saturated. In this case, we estimated the encircled energy on the unsaturated companion star and scaled this with the measured contrast on unsaturated images. The Strehl ratios for the various tests we conducted are listed in Table \ref{tab:observations}, and the integrated PSFs for a subset of the tests are shown in Fig. \ref{fig:psfs}. We note that our method for estimating the Strehl ratio likely slightly overestimates it, as it does not consider the halo outside 30 $\lambda/D$. Still, it allows us to compare the performance of our approach with the standard modulated PWFS.

We obtained a slightly lower Strehl ratio with the smart unmodulated PWFS than with the modulated PWFS for observations of  $\alpha$ Eri and $\pi$ Pup. The main limitation for both reconstruction methods in good seeing conditions, as was the case for the $\alpha$ Eri observations, is residual tip-tilt vibrations. On the other hand, we observe an improvement by a factor two for the Strehl ratio in observations of AF Lep. This is a fainter star, and these observations suffered from strong winds. This improvement could be the result of the improved sensitivity of the unmodulated PWFS and/or the better nonlinear estimation that we have with the CNN. From the wavefront sensor images, we estimated a signal-to-noise ratio of about $\sim6$ per pixel in the pupils of the PWFS for the AF Lep observations. Tests with the CNN were conducted before and after the tests with the modulated PWFS; the performance was  similar, so it is unlikely that the improved Strehl ratio is the result of changing atmospheric conditions. This is an encouraging result, and the performance on fainter stars will be tested in more detail in the future. 

A lower Strehl ratio was obtained for the November 20, 2024, tests compared to the modulated PWFS. This may be the result of a slightly worse model due to, for example, drifts during the data collection, but may also be the result of worse conditions leading to larger nonlinearities that are still hard to correct with the unmodulated PWFS. Training on more data should help improve the performance in these conditions. We also attempted to run the model at 3.6 kHz to see the limits of the current implementation. Unfortunately, we observed that every $\sim$9 frames of the science camera, or every $\sim$1.1 seconds, there is a frame with a much lower Strehl ratio. This pattern is regular and is likely due to jitter in the latency. We did not see this in the isolated latency tests presented in Table \ref{tab:latency}. This means that the jitter is not due to TensorRT but is likely due to interaction with other processes running on the MagAO-X real time computer. Running the model on a dedicated GPU without other tasks may mitigate this latency issue. Additionally, using the model at half precision may help reduce the latency and make speeds of $>$3 kHz feasible, with a minimal impact on the reconstruction accuracy \citep{Pou2024SPIE_real_time_cosmic}. Alternatively, the jitter may be the result of an issue with the DM kernel module, which was found to have some problems when running at $\gtrsim2.5$ kHz.

\begin{figure}[htbp]
    \centering
    \includegraphics[width=\linewidth]{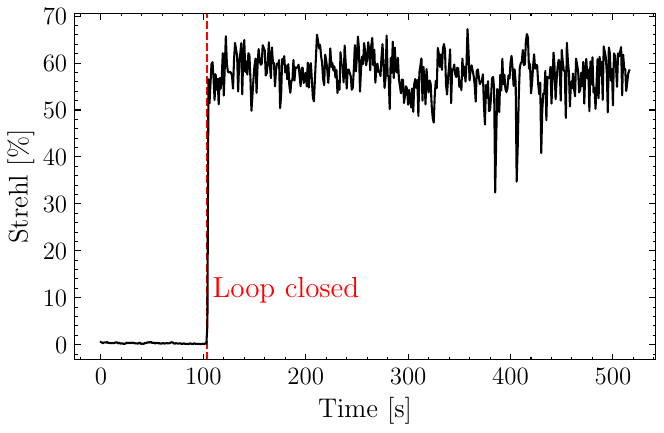}
    \caption{Measured Strehl ratio as a function of time for the test on $\alpha$ Eri on November 16, 2024, showing a generally stable performance over time. The vertical dashed red line indicates the time at which the control loop was closed. There are three frames in which the Strehl ratio drops significantly due to latency on the GPU.}
    \label{fig:strehl_time}
\end{figure}

Figure \ref{fig:strehl_time} shows the evolution of the Strehl ratio over time for the $\alpha$ Eri observations. We see that the performance was generally stable over time once the loop was closed. However, we observe that for three frames the Strehl ratio drops to 40\% or lower. Upon inspection of these images, we conclude that this is also likely the result of a jitter in the latency of the inference. While it is much less common at 2 kHz than at 3.6 kHz, running the model on a dedicated GPU and going to a half precision model should also mitigate this issue at 2 kHz. Outside these frames, the performance appears to be stable, and we do not observe, for example, the accumulation of specific modes on the DM.

\section{Conclusions and outlook}\label{sec:conclusions}
We have presented the first on-sky demonstration of XAO with a nonlinear NN reconstructor for an unmodulated PWFS. We have shown that the performance with this combination can approach that of a highly tuned linear reconstruction with the modulated PWFS on bright stars and under good conditions. Additionally, the Strehl ratio was higher for observations of a slightly fainter star under the influence of strong winds. Our results show that using the PWFS in its unmodulated, and most sensitive, form is feasible when using a nonlinear reconstructor. This improved sensitivity will help improve the performance of MagAO-X and other XAO systems on fainter stars \citep{Males2016SPIE.9909E..52M, Chambouleyron2023A&A_noise_propagation_ffwfs, Agapito2023_nonmodulated_pwfs}. Furthermore, eliminating the need for modulation can simplify instrument designs and remove a point of failure. Additionally, it will allow XAO systems to run the control loop at faster speeds on bright stars and improve the contrast at small angular separations. This is due to increased sensitivity and the absence of the need for modulators operating at very high frequencies. Finally, the unmodulated PWFS is sensitive to petal-piston and/or segmenting errors \citep{Bertrou-Cantou2022_petalpiston_pwfs, Hedglen2022_segment_phasing_pwfs, Engler2022_flipflop}, allowing the XAO system to control those modes at high speeds. The ability of NNs to reconstruct these modes with the unmodulated PWFS will be studied in future work. \\

We plan on conducting additional tests during upcoming observing runs. First, we want to test the performance on faint stars and see if we can push the limiting magnitude at which we can use XAO. This will require training the model on the appropriate noise levels. Additionally, we want to implement the model at half precision on a dedicated GPU in order to run the control loop at faster speeds ($>3$ kHz) and decrease the temporal error. Furthermore, we hope to improve the calibration procedure by moving the data collection from a Python script to C++. This will reduce the time it takes to collect the calibration data and should help prevent slow drifts in the instrument from impacting the training data. Collecting the full training dataset should only take 100 seconds when running at 1 kHZ. Alternatively, the loop can be closed at certain points during the data collection to ensure we are calibrating around a flat wavefront.

Finally, we plan on studying the limits of using the smart PWFS for XAO on the ELTs in the context of the Planetary Science Camera (PCS; \citealt{kasper2021pcs}) and Giant Magellan Adaptive Optics eXtreme (GMagAO-X) instruments \citep{Males2024_gmagaox, close2024_gmagaox2024, haffert2024_gmagaox}. Initial tests of the latency for ELT-sized systems are presented in Appendix \ref{app:A} and show that it is feasible to use our approach for XAO on ELTs.

\begin{acknowledgements}
We thank the anonymous referee for the comments that have improved the quality of this work. The MagAO-X phase II project acknowledges generous support from the Heising-Simons Foundation. We are very grateful for support from the NSF MRI Award \#1625441 (MagAO-X).
\end{acknowledgements}

\bibliographystyle{aa}
\bibliography{references}

\begin{appendix}
\section{Latency for ELT-sized systems}\label{app:A}
\begin{figure}[htbp]
    \centering
    \includegraphics[width=\linewidth]{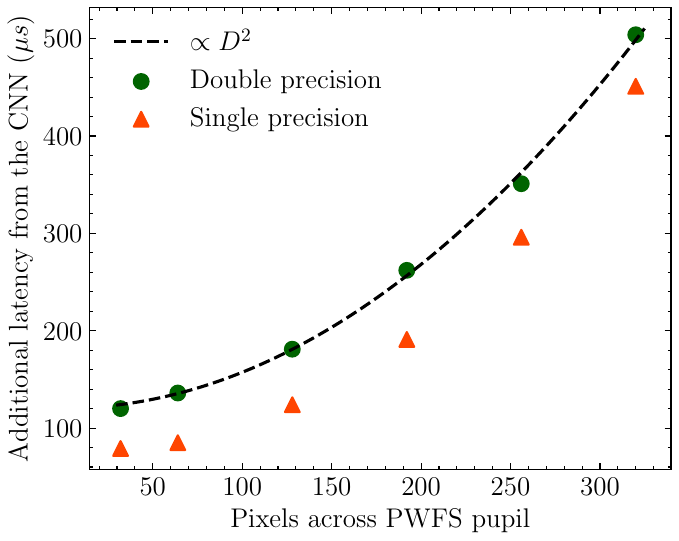}
    \caption{Latency of the convolutional part of the CNN as a function of the number of pixels across the PWFS pupil using the MagAO-X GPU. For small telescopes this is dominated by overheads, while for large telescopes we find a quadratic relationship between latency and telescope diameter.}
    \label{fig:latency_scaling}
\end{figure}

To test the feasibility of using our approach for ELT-sized systems, we tested the latency of the convolutional part of our NN for increasing number of pixels across the PWFS pupil. We ignore the linear part of the model as that is required for both the CNN and the standard MVM reconstruction. The results of the latency tests are shown in Fig. \ref{fig:latency_scaling}. We find that for MagAO-X we are dominated by overheads such as memory transfer. In the limit for a large number of pixels there is a roughly linear relation between the latency and the total number of pixels in the wavefront sensor. This is equivalent to a $\propto D^2$ relation with $D$ the diameter of the telescope. On the other hand, the number of operations for the MVM is proportional to both the total number of pixels in the wavefront sensor and the number of corrected modes, combining to a roughly $\propto D^4$ relation when not using sparse matrices. We note that this assumes that we do not need to use a deeper model with more layers or channels to obtain the same reconstruction accuracy for the CNN, which needs to be verified. For both approaches, a distributed computing approach might help ease these scaling relations. Additionally, the compute capabilities of GPUs has rapidly improved over the years and should continue to do so. This prospect, together with the measured latencies for ELT-sized systems in Fig. \ref{fig:latency_scaling}, gives us confidence that using CNNs at kilohertz frequencies for the ELTs will be feasible.

\end{appendix}

\end{document}